# Evolution of the genetic code.
# From the CG- to the CGUA-alphabet,
# from RNA double helix to DNA.

Semenov D.A. (dasem@mail.ru)
International Research Center for Studies of Extreme States of the Organism at the Presidium of the Krasnoyarsk Research Center, Siberian Branch of the Russian Academy of Sciences

**Abstract.** *A hypothesis of the evolution of the genetic code is proposed, the leading mechanism of which is the nucleotide spontaneous damage leading to AT-enrichment of the genome. The hypothesis accounts for stability of the genetic code towards point mutations, the presence of code dialects, emergence of stop codons, emergence of the DNA double helix and the symmetry of the genetic code table. The assumption of the originally triplet structure of the genetic code has been substantiated. A hypothesis concerning the primary structure of the first gene and the first protein has been proposed.*

**Introduction.**
The history of the studies of the genetic code evolution began with the publication of Nirenberg's dictionary in 1966 [1]. That same year saw the publication of the first study in this area [2]. Rumer's pioneering study of the structure of the genetic code was published in 1966, too [3].
These two currents first met in the classic paper by F. Crick "Origin of the genetic code" in 1968 [4]. The author proposed the "frozen accident" hypothesis and, at the same time, demonstrated the structural order in the genetic code that was different from the symmetry discovered by Rumer. This paper has largely determined the attitude of biologists towards the issue of the genetic code evolution.
Chronologically, the first were Rumer [3] and Crick [4] with their studies on the structure of the table of the genetic code. Then, in 1975, Hartman suggested that nucleotides were incorporated into RNA in a certain order [5] and in 1978 he specifically assumed that cytosine and guanine had been before uracil and adenine [6]. Hartman tried to prove that statement in his later studies, based on the evolution of tRNA and aminoacyl-tRNA-synthetases [7, 8]. As I adhere to the idea that just guanine and cytosine existed at the initial stage of the evolution of the genetic code, it makes no sense to compare structures of different tRNA and aminoacyl tRNA synthetases.
Let's imagine that one studies radio signals and responses to them in a complex system. It would be logical to assume that similar signals should cause similar responses. This is rational in terms of language structure. Signals are received by radio receivers. Does it make sense to make a detailed comparative analysis of the structures of radio receivers? The structure obviously contains information, but is it all useful? I only use the information on the structure of the anticodon. The codon (signal) changes gradually, properties of encoded amino acids (response) change gradually, the anticodon (some definitely significant part of the receiver) changes gradually.
Although agreeing with Hartman's statement above, I have to reject his assumption that adenine emerged before uracil. This could not be true because adenine can only be complementary to uracil and, moreover, there is a mechanism of spontaneous conversion of cytosine to uracil while there is no mechanism of spontaneous emergence of adenine.

In 1993, Hornos's study [9] opened a new stage in the investigation of the structure of the genetic code. The main achievement of the author, as I see it, was not the use of the contemporary mathematical tools, but rather the attempt to establish an immediate relationship between the structure and the evolution of the genetic code. The next appreciable step was, in my opinion, a study performed by Bushford [10], in which both the table of the genetic code and the table of anticodons were used as information sources.

The structure of the genetic code was shown to be related to its evolution in papers by Hurst [11,12]. Among other things, the author demonstrated that the universal genetic code is amazingly more stable to point mutations than random variations of genetic codes.

The study conducted by Knight [13], which demonstrated the unique ability of arginine to directly recognize its codon, is in good agreement with my hypothesis, too.

Jordan and co-authors [14] related the discovered trend of amino acid gain and loss by proteins to the process of the genetic code evolution. However, it would be more correct to treat this trend as ensuing from the structure of the genetic code and the presence of spontaneous single mutations. As will be proved below, these two events themselves are closely related to the evolution of the genetic code.

None of the studies mentioned above proposes a physicochemical mechanism of the emergence of new nucleotides. Also, most of the authors consider the "evolution of the genetic code" as an abstract process of filling the already existing table. However, the process that leads to the conversion of cytosine to uracil has been thoroughly investigated and is well known – this is spontaneous deamination of cytosine. Cytosine is unstable and is, at a certain rate, deaminated, which leads to accumulation of uracil.

The sequentially arising instabilities induce the emergence of adenine and the first elements of grammar in the genetic code – stop codons. These instabilities are caused by physical processes and their emergence can be studied experimentally.

From a certain moment on, guanine can be replaced by uracil in the course of the oxidative guanine damage. This process is also spontaneous, but it cannot occur in the absence of adenine and the DNA double helix.

Two thoroughly investigated typical mutations: cytosine deamination and the oxidative guanine damage have become a source of diversity in codon sequences. The new codons that are insignificantly different structurally will correspond to the new amino acids whose properties are similar to those of their precursors. Thus, the notion of evolution as development through small changes, which is accepted in biology, can be extended to the evolution of the genetic code. So, to "derive" the table of the universal genetic code and any of its dialects, one only needs to know the initial sequence of codons and the properties of their respective amino acids.

**Emergence of new nucleotides and new amino acids .** To account for the extant structure of the genetic code (all dialects), I can propose the hypothesis of the evolution of the current four-letter alphabet from an earlier, two-letter one. In this hypothesis, deamination of cytosine plays the key role. Remembering that DNA and cytosine methylation seem to be of quite recent origin, it would be correct to write C→U.

Cytosine deamination obviously causes just partial loss of complementarity. There can be two hydrogen bonds between guanine and uracil, at first glance making this pair similar to the adenine-thymine (uracil) pair. The possibility of the formation of the complementary pair is illustrated by coupling of these nucleotides at the third position of the codon and anticodon [15] (Table 1). This ambiguous complementarity can successfully be accounted for by the stabilization of uracil in the enol form in the G-U pair [16], which will be discussed below.

If RNA replication involves the products of hydrolysis of the earlier RNA molecules, uracil will be among disintegration products and will be incorporated into new copies. If the copying accuracy is low, uracil fraction will grow quite rapidly.

| anticodon | codon |
|---|---|
| C | G |
| G | C, U |
| U | G, A |
| A | U |

Table 1. Ambiguity of nucleotide coupling at the third position of the codon and anticodon [15].

Based on the assumption that the two most complex nucleotides, guanine and cytosine, were before uracil and adenine, one can reconstruct the stages of the genetic code evolution.

Table 1 clearly illustrates that if there were just two letters (C, G), the possibility of the formation of the G=U pair facilitated the process of incorporating the new letter into the code, and in the next stage, the possibility of the formation of the U=A pair facilitated the emergence of adenine. Formally speaking, a reverse order of the emergence of these letters is also possible – first, (A, U), then, G and, last, C. The realization of this order, however, is physically ungrounded: with every step, the letters would have to get more complex. The evolution from (CG) to (CGUA) was propelled by spontaneous mutations: cytosine deamination and the oxidative guanine damage [17]. This is the so-called process of the genome AT-enrichment. One can say that the evolution of the code was thermodynamically determined.

Long chains consisting of guanine and cytosine could encode the first four amino acids: proline, glycine, alanine, and arginine. In all dialects of the genetic code, these amino acids are encoded by the same letter doublets (i.e. the first two nucleotides of the triplet).

Deamination led to gradual accumulation of uracil, which was initially read as cytosine.

Due to accumulation of a considerable amount of uracil the new letter doublets – CU, UC, and GU – acquired meaning. The presence of similar amino acids that correspond to these strong letter doublets (encoding only one amino acid) in all dialects is indicative of evolutionary antiquity of the letter doublets.

Let us consider the table of the genetic code letter doublets as it was proposed by Yu.B. Rumer in 1968 [3, 18] (Table 2). The author focused his attention on the presence of the letter doublets (i.e. the first two nucleotides of the triplet, called "roots" by Yu.B. Rumer) and their ability or inability to encode just one amino acid. Of 16 letter doublets, 8 were strong (encoding just one amino acid) and 8 were weak (encoding more than one amino acid). Color gradation illustrates the order of filling of the table cells: the darkest cells correspond to the most ancient letter doublets.

The table of the genetic code letter doublets should be filled starting with the four earliest and strongest letter doublets and moving towards the four weakest and latest.

|   | C | G | U | A |
|---|---|---|---|---|
| C | Pro | Arg | Leu | His / Gln |
| G | Ala | Gly | Val | Asp / Glu |
| U | Ser | Cys / Trp/Stop | Phe / Leu | Tyr / Stop |
| A | Thr | Ser / Arg | Ile / Met | Asn / Lys |

Table 2. The order of filling of the cells in the table of genetic code letter doublets

Thus, we get Pro→Ser, Pro→Leu and Ala→Val. The close relationship between proline and serine looks odd, but in proteins proline is often present as hydroxyproline, and hydroxyproline is similar to serine. Filling of new table cells by amino acids similar to the preceding ones is a basis for the observed stability of the genetic code towards point mutations, primarily mutations leading to the genome AT-enrichment.

At that stage, the UU letter doublet did not acquire meaning because it was rare: the amount of uracil was still low.

The UG letter doublet could have acquired meaning later than CU, UC, and GU, which has to be additionally accounted for and will be discussed in detail in the chapter about emergence of stop codons.

The last step in the formation of the alphabet was incorporation of adenine. The uracil-guanine pair must have distorted the complementarity of the neighboring nucleotides and must have been insufficiently strong.

Selection resulted in the emergence of adenine as a better pair for uracil. The relatively late emergence of adenine is indicated by the fact that differences in dialects of the code are localized in the letter doublets involving adenine, except the UG and UC letter doublets [19].

With the emergence of adenine, a new letter doublet could arise not only due to cytosine deamination but also as a result of the oxidative guanine damage.

The most probable order of the filing of cells is:

**Pro→Leu→Leu(Ile)(Phe); Pro→Ser(UC)→Tyr; Pro→Thr→Met**
**Arg→Arg→Lys; Arg→His/Gln→Asn; Arg→Trp**
**Ala→Val→Ile(Phe)**
**Gly→Asp/Glu; Gly→Ser(AG); Gly→Cys→Tyr**

The most convincing illustration of the proposed mechanism of the genetic code evolution is the fact that all amino acids similar in properties to arginine have their origin in the arginine letter doublet, CG.

Glycine seems to be different in its chemical properties from the amino acids that emerge as a result of mutations of its codons, but these amino acids (Cys, Ser, Asp, Glu) are all alike, rather suggesting an incompleteness of our knowledge regarding similarities in the properties of amino acid molecules.

This similarity in chemical properties does not allow us to unequivocally determine the precursor amino acids for isoleucine (Ile), tyrosine (Tyr), and phenylalanine (Phe).

To solve this problem for the tyrosine amino acid, let us trace the mutations that result in filling of other cells of the table. Cytosine deamination in the first position of the codon will be denoted as C1 and in the second – C2; the oxidative guanine damage in the first position of the codon will be denoted as G1 and in the second – G2. I apologize to mathematicians for introducing these notations. For such operations in the complementary chain leading to the emergence of adenine in the codons I introduce underlining. For isoleucine and phenylalanine, both variants of their possible origin are given. Results are presented in Table 3.

Please note that there are no mutations G2 and G2 in the list, but if we allowed **Ser→Tyr**, we would have to allow these mutations, which would be groundless. Thus, the most probable direction of populating the table is **Cys→Tyr**.

| Pro | C2 | Leu | C1 | Leu |
|---|---|---|---|---|
| Pro | C1 | Ser | | |
| Pro | G1 | Thr | C2 | Met |
| Arg | G1 | Arg | C2 | Lys |
| Arg | C2 | His/Gln | G1 | Asn |
| Arg | C1 | Trp | | |
| Ala | C2 | Val | C1 | Ile |
| Val | G1 | Phe | | |
| Leu | C1 | Phe | | |
| Leu | G1 | Ile | | |
| Gly | G1 | Cys | | |
| Gly | C1 | Ser | | |

Table 3. Analysis of mutations leading to the filling of new cells.

**Symmetry of the genetic code table.** Let us consider some specific features of the symmetry of the genetic code, again turning to the table of the codon letter doublets. Rumer proposed this representation of the table of letter doublets for the genetic code in order to illustrate the symmetry he had discovered [3,18] (see Table 4).

|   | **C** | **G** | **U** | **A** |
|---|---|---|---|---|
| **C** | Pro | Arg | Leu | His / Gln |
| **G** | Ala | Gly | Val | Asp / Glu |
| **U** | Ser | Cys / Trp/Stop | Phe / Leu | Tyr / Stop |
| **A** | Thr | Ser / Arg | Ile / Met | Asn / Lys |

Table 4. The symmetry of the table of the letter doublets for the genetic code (according to Rumer). Strong letter doublets are marked in gray.

Examination of the symmetry of the genetic code table seldom involves use of Rumer's canonical sequence: C>G>U>A [18]. Much more frequently used is Crick's sequence: U>C>A>G [4], and the table presented in the circular form. The two different ways to arrange nucleotides in sequences highlight different properties of the code. Rumer's sequence indicates the number of hydrogen bonds in complementary pairs and Crick's sequence – the indistinguishability of A and G (C and U).

| AU ↑ | CU → ↑ | UU | | AU | CU | UU |
|---|---|---|---|---|---|---|
| AC ← | CC → | UC | | AG ← | CG → | UG |
| AA | CA | UA | | AA ← | ↓ CA | × UA |

| AU | GU | UU | | AU ← | GU → | UU |
|---|---|---|---|---|---|---|
| AG ← | GG → | UG | | AC | GC | × UC |
| AA | ↓ GA | ↓ UA | | AA | GA | UA |

Table 5. A schematic representation of filling of the table of codon letter doublets.

Please note that this study does not give any explanation to Rumer's symmetry. It is used to emphasize the natural direction in which the cells are filled. If the cells were filled from the weak letter doublets to strong ones, the genealogy of codons would not be unambiguous. The symmetry reflected in the circular genetic code table (Crick's symmetry) can be entirely explained by the proposed hypothesis. The circular table demonstrates stability of the genetic code towards cytosine deamination in the second and the first positions of the codon and its stability towards oxidative guanine damage only in the first position of the codon.

Table 5 presents the variant of the most likely filling of the cells in the table of the genetic code letter doublets, showing their relationships. The absence of the G2 mutation is evident: the downward movement from the CC and GC letter doublets and the upward movement from the GG and CG letter doublets are impossible. In Table 4 this corresponds to codons' kinship in either odd or even columns. This determines the symmetry of the genetic code.

**Dialects of the genetic code.** Let us now consider the differences characteristic of various dialects of the genetic code [19]. Differences relating to stop codons will not be discussed. Table 6 presents variants of filling for different dialects of the code compared with the universal genetic code.

The other existing variants of the codes differ just quantitatively, i.e. by the number of codons corresponding to the amino acid in a given cell rather than by the amino acids themselves. This variant is given in the second line of the table to demonstrate that Rumer's symmetry is not universal either.

Based on the presence of threonine in the CU letter doublet in the genetic code of yeasts, we should give preference to the origin of isoleucine from valine: Val→Ile. Moreover, the GU letter doublet corresponding to valine and the AU letter doublet corresponding to isoleucine are related by their ability to encode the initiation codons.

The proposed hypothesis accounts for stability of the genetic code towards point mutations, the presence of code dialects, and the symmetry of the genetic code table.

| Paramecium, Tetrahymena, Oxytrichia, Stylonychia, Glaucoma, Acetabularia | UAA, UAG—Gln instead of Stop | Gln(CA) →Gln(UA) |
|---|---|---|
| Molluscs, Echinoderms, Platyhelminths, Nematodes | AGA, AGG—Ser instead of Arg | Gly→Ser |
| Yeasts | CUN—Thr instead of Leu | Pro→Thr |
| Candida cylindrica | CUG—Ser instead of Leu | Pro→Ser |
| Ascidians | AGA, AGG—Gly instead of Arg | Gly(GG)→Gly(AG) |

Table 6. Significant differences in the existing variants of the genetic code dialects [19].

**Emergence of stop codons.**

In the first part I wrote that the UG letter doublet, unlike GU, CU and UC, acquired its meaning after the emergence of adenine. Thus, in my opinion, this letter doublet is the first version of a stop codon. In the table of the universal genetic code, this letter doublet is a stop codon; moreover, this letter doublet encodes cysteine and tryptophan.

What could cause the emergence of non-encoding codons in the course of evolution of the genetic code? I would venture to suppose that the reason was quite simple – relative instability of complementary interaction of the codon with the respective anticodon or, more exactly, with the part complementary to the letter doublet.

The UG letter doublet could be essentially different from the GU pair in its interaction with the complementary dinucleotide. As we are considering the stage at which there was no adenine yet, we should investigate the interaction of the UG and GC dinucleotides.

There are reliable data that can provide a basis for substantiating my hypothesis prior to experiments and calculations. Let us examine the table of the anticodon-codon correspondence at the third position (Table 1).

Uracil can complementarily interact with adenine and guanine, but at the first and second positions of the codon there are always classic uracil-adenine pairs, indicating greater energy efficiency of this interaction. If there were no energy benefits, there would be frequent amino acid replacements in proteins, due to a foreign anticodon joining the codon.

Thus, we can make the conclusion that the uracil-guanine pair was less stable than the uracil-adenine one. This alone reduces the stability of complementary interactions of all letter doublets that originated at the stage of uracil incorporation into the genetic code.

Let us now discuss how the rearrangement of letters in the pair could affect the stability of complementary interaction. The DNA double helix is a system with very few degrees of freedom: the monomers are joined together not only by phosphodiester bonds but also by numerous hydrogen bonds. The two hydrogen bonds in the adenine-thymine pair strongly restrict the movement of nucleotides relative to each other and the three bonds in the guanine-cytosine pair make conformational changes just impossible. It is amazing that DNA can retain complementarity along its entire length: the existence of such a structure without conformational strain is highly unlikely. DNA is certainly a product of evolution and the ability of the double helix to exist without strain is a selection factor. It has been generally accepted that the presence of deoxyribose instead of ribose and thymine instead of uracil in the DNA structure is somehow related to the stability of the double helix.

The most likely carriers of genetic information in the early stage that we are discussing were RNA molecules; thus, it seems appropriate in this case to use the data on RNA oligonucleotide interactions. The unrestricted RNA molecule cannot form a double helix, i.e. it is spatially (sterically) more hindered than the DNA molecule. Replacements of nucleotides in the RNA structure could significantly affect the stability of their interactions with the complementary oligonucleotides.

We can obtain the necessary information on RNA oligonucleotide interactions by considering the correspondence between codons and anticodons. The uracil-guanine pair is sometimes formed at the third position, during codon-anticodon interaction. Let us insert the data on the events of the emergence of this pair into the table of the genetic code letter doublets [10, 15] (Table 7).

|   | C | G | U | A |
|---|---|---|---|---|
| **C** | I | I | I | G |
| **G** | I | G | I | G |
| **U** | I | G | G | G |
| **A** | I | G | I | G |

Table 7. Nucleotides complementary to uracil at the third position of the respective codons (eukaryotic genetic code). Strong letter doublets are marked in gray.

For eight letter doublets the third position of the anticodon is occupied by guanine and for the other eight – by inosine. Please note that the distribution of anticodon endings is almost the same as the distribution of the strong and weak letter doublets. The presence of guanine or inosine at the third position can only be accounted for by the energy benefits associated with the presence of a given nucleotide in given surroundings. All letter doublets for which inosine at the third position of the anticodon is advantageous must be of a similar shape, which determines higher efficiency of inosine compared to guanine.

As evident from Table 7, the presence of inosine at the third position of the anticodon is strongly dependent upon the second letter of the codon. The presence of pyrimidine at the second position of the codon leads to the presence of inosine in 7 out of 8 cases. The rearrangement of purine and pyrimidine in the codon letter doublet (except GC) causes a change in the shape of the codon and, as a result, in the anticodon ending. As the purine and the pyrimidine bases significantly differ in their sizes, the above statement seem nearly obvious.

The considerable changes in the shape of the codon may account for the impossibility of evolutionary filling of the cells in the genetic code table as a result of the G2 mutation (the oxidative guanine damage in the second letter of the codon). This mutation leads to the replacement of purine by pyrimidine, significantly changing the shape of the codon if the damage occurs in the second letter. Table 7 shows that the oxidative guanine damage in the first letter is not so significant.

Please note that the UG letter doublet differs from CU, UC and GU by the presence of purine (guanine) at the second position and guanine at the third position of the anticodon, and, thus, has a different shape.

The UG letter doublet must have been much more weakly bound to complementary dinucleotides, which provided the basis for the emergence of the first stop codons. I'd like to point out that here I present a physical basis for the existence of termination signs in the genetic code. This is a greater event than the emergence of a new amino acid – the emergence of stop codons must have fundamentally changed the arrangement of genetic texts. When there were only cytosine and guanine, the peptide length could not be regulated, which limited the functional capabilities of proteins. One can say that stop codons brought the opportunity to diversify the meanings of protein sequences.

If the analogy with the evolution of human language is relevant here, the emergence of stop codons can be compared to the transition from phonemes to sentences. Before stop codons emerged, the genetic code had no grammar and there were no rules of handling words. The impossibility to form a complementary structure (due to its instability) can be regarded as a serious problem and inconvenience. However, this inconvenience was transformed into a great finding in the course of the genetic code evolution. That was how the "baby-talk" of the first proteins developed into the "meaningful phrases" of modern design.

The emergence of adenine makes the UG letter doublet much more stable, and this must cause it to lose its natural properties allowing it to be a stop codon. The letter doublets consisting of uracil and adenine only are now more suitable for this role. This function is inherited by the UA letter doublet (for the universal genetic code). Now the UG letter doublet can be occupied by amino acids. In some dialects of the genetic code the UG letter doublet can even lose its initial function entirely and then all of its codons encode cysteine and tryptophan.

For the function of the stop codon to stabilize, there should have been a reasonable time interval between the incorporation of uracil and adenine into the code. It seems that this time interval lasted almost until the formation of the DNA double helix.

**Emergence of DNA.**
The hypothesis that at the early stages there were just cytosine and guanine makes the picture of the RNA world much poorer. It is difficult to form a great variety of tertiary structures based on nucleotides that tend to form almost exclusively Watson-Crick's pairs. This hypothesis can be said to restrict and, thus, to simplify the picture of the early RNA world.

However, this hypothesis can provide an opportunity to trace logically the process of the emergence of the DNA double helix. The generally accepted picture of the RNA world suggests that DNA was invented to separate the function of information storage from the function of translation. The double copy is more stable towards interferences and mutations, but less accessible to enzymes. The emergence of DNA is often associated with the genome increase, suggesting that smaller organisms such as viruses can make do with RNA. This interpretation ignores one low-probability assumption – it is accepted de fide that a ***random*** replacement of ribose by deoxyribose resulted in the transformation of the previously single-strand RNA into the double helix. To make the copy more stable towards mutations! But this is pure teleology!

Yet, nucleotides themselves already possess complementarity properties that allow them (in a miraculous way) to be incorporated into this double helix. If this did not occur at the early stages and complementarity was necessary for the nucleotide-to-nucleotide replication, how can one account for this amazing correspondence?

The hypothesis suggesting that at early stages of the evolution of the genetic code there were just cytosine and guanine makes it possible to bring the emergence of DNA back to evolutionary philosophy.

The complementary long chains based on guanine and cytosine only (CG-polyribonucleotides) are more apt to form the double helix than the complementary polyribonucleotides based on all four pairs. The reason is simple: guanine and cytosine form just Watson-Crick's pairs due to a significant energy benefit as compared to other variants.

I can admit that different variants of oligomeric complementary RNAs based on cytosine and guanine can yield double helixes of different stabilities. It is important that for a random sequence of CG ribo-oligomer the RNA double helix will be stable.

The next step in my investigation will be emergence of uracil due to cytosine deamination. Deamination first leads to the formation of uracil in the enol form. Then it must be transformed into the keto form, which is more stable in solution. However, when uracil is gripped in the backbone of the RNA double helix, the enol form is more advantageous for it. Uracil in the enol form can make two and even three hydrogen bonds with guanine, while uracil in the keto form would be able to make only one hydrogen bond. One can say that the enol form of uracil is stabilized by its interaction with the complementary guanine. The keto-enol tautomerism is a well-studied process and chemists will understand me without any further proofs. I should add that Watson first considered nucleotide formulae in the enol form as the more probable, guided by reference books of that time [20]. Fortunately, I can give convincing arguments using commonly available information. Codon-anticodon interaction results in the formation of a short segment of the double helix. In the case of GC-rich codons, opposite to uracil in the third position of the codon there is inosine in the anticodone.

Crick's wobble hypothesis [21] allows a solution for this pair only by wobbling the third nucleotides of the codon and the anticodon. However, in the double helix, their position is also stabilized by stacking. If stacking exerts significant influence so that wobbling becomes impossible, then no uracil-inosine complementarity is possible for the keto form of uracil, i.e. there cannot be any hydrogen bonds. In this case, uracil is in the state that most closely imitates cytosine.

Although Crick's idea is certainly true and proved to be fruitful, it is superfluous for explaining the codon multivariant pairing [16]. If C-I are Watson-Crick's pairs and U-I pairs are formed in accordance with the wobble hypothesis, it is not clear why these codons are always indistinguishable, although they have different conformations, which can be stabilized. With the keto-enol tautomerism, both codons have the same conformation.

Why is uracil opposed by inosine is some anticodons and by guanine in others? According to Crick's hypothesis inosine and guanine are indistinguishable to uracil. If we suppose the presence of the enol form of uracil, we can suggest that inosine emerged in anticodons in an evolutionary way because in this case the guanine amino group could not form a hydrogen bond. That is, in this case inosine (guanine without an amino group) is sufficient. I should add that the enol form of uracil can be registered in NMR spectra, which makes the necessary experiments easy to perform.

Uracil incorporated into the RNA double helix is very similar to cytosine. Accumulation of the sufficient amount of uracil gradually makes the double helix unstable as the bond is not so strong. The loss of stability at a certain section is accompanied by the transformation of uracil into the keto form and breakdown of the double helix. It is the occurrence of this instability (catastrophe) that causes the system to become more complex. For uracil to continue accumulating in the double helix, the RNA double helix must grow more stable. The greater stability is attained with the "invention" of adenine, which can form two hydrogen bonds with uracil in the keto form.

The emergence of adenine is a source of strain in the ribophosphate backbone of the RNA double helix. In AT-rich codons, uracil in the third position of the codon is opposed by guanine in the third position of the anticodon, suggesting a distortion in the codon conformation.

Accumulation of AU pairs leads to the already familiar catastrophe – loss of double helix stability. The way out of this catastrophe cannot be achieved by conventional means but only by the system becoming more complicated. The replacement of ribose by deoxyribose reduces the rigidity of the sugar-phosphate backbone. Thus, DNA emerges.

Uracil methylation and emergence of thymine must have been associated with the stage at which such a catastrophe occurred. Methylation, e.g., could favor further stabilization of uracil in the keto form. Oxygen is an electron acceptor while the methyl group an electron donor, so the presence of the methyl group reduces the probability of the proton being near oxygen.

**Strong and weak letter doublets. Rumer's symmetry.**

The keto-enol tautomerism is a good example of dramatic changes in the form and properties of certain nucleotides that occur under rather weak interactions. While one form of the sugar-phosphate backbone is favorable to the enol form of uracil, the other is favorable to its keto form.

A nucleotide triplet can be also assumed to change its form, but in this case a change in the form of the first two nucleotides occurs due to the third. Here I do not speak of the form of the triplet in the solution or the form of the triplet in mRNA, but rather the form of the double helix segment resulting from the codon-anticodon interaction.

This form is determined by two forces – complementary interactions and stacking (interaction of neighboring nucleotides). Stacking is a nonspecific event as there is a well-known formula: purine-purine>purine-pyrimidine>pyrimidine-pyrimidine.

In the third position of the codon the number of the formed hydrogen bonds is much less relevant than the nature of the bases – purine or pyrimidine. Thus, the presence of strong and weak letter doublets (Table 4) can naturally be related to the presence of stacking.

In codons with the CC, CG, GC, and GG letter doublets, the form of the doublet is only determined by complementary interaction. The three hydrogen bonds in each doublet make conformational changes impossible.

Let us now discuss the (UC-UG), (AC-AG), (CU-CA) and (GU-GA) letter doublets. In each doublet the first letter is strong and the second weak. The number of hydrogen bonds in each doublet is the same, but doublets with purine in the second position are the weak ones. Due to the presence of purine in the second position, two purines can occur one by one – in the second and the third positions of the codon. This construction imparts sufficient stress to the first two nucleotides to cause a change in their conformation. Please note that the anticodon is much less conformationally flexible because it is part of tRNA and is stabilized by its structure.

In codons with the UA and AA doublets, the conformation of the doublet can be changed due to the presence of purine in the third position of the codon. In the codons with the UU and AU doublets, the conformation of the codon can be changed by just a slight interaction between pyrimidine and purine.

The fact that complementary interactions are comparable with stacking in their effect may be surprising, but if this were not so and there were just one predominant type of nucleotide interactions, the other would not be described in textbooks. It is not less surprising that only half of the codon letter doublets are capable of conformational changes under the impact of stacking. This may be a significant fact, but I have no ideas concerning this.

Rumer's symmetry [3, 18] can then be interpreted as follows: U>A because pyrimidine in the second letter of the codon prevents it from possible stacking-related conformational changes.

In the table of the universal genetic code there are two amino acids, each of which is encoded by one codon only: tryptophan (Trp), encoded by the UGG codon, and methionine (Met), encoded by AUG. This feature can be consistently explained based upon the same arguments that have been used to explain Rumer's symmetry. If there are letter doublets incapable of conformational changes when pyrimidine is replaced by purine in the last letter and there are letter doublets that readily change under these conditions, there must be letter doublets that are close to the equilibrium point. For these letter doublets even much smaller impacts may prove to be significant. The UG doublet is situated on the table diagonal, i.e. it may be close to the equilibrium point. For its conformational change it may be important not only that the third position is occupied by purine but also that this is guanine. An additional hydrogen bond leads to a conformational change. The situation with the AU doublet is similar, but instead of the purine-purine interaction, we have to assume that its conformation can be changed by a weaker, pyrimidine-purine, interaction. In this case, the letter doublet itself is extremely prone to conformational changes: the doublet contains no nucleotides capable of forming three hydrogen bonds. The AU doublet is presumably close to the equilibrium point and the conformational transition is achieved at the expense of one hydrogen bond in the last nucleotide.

Thus, it is the first two nucleotides, or, to be more exact, the conformation of the first two nucleotides, that are the encoding part of the codon. For half of the letter doublets this conformation depends upon the last nucleotide, although it is this conformation rather than the whole sequence that is recognized. The number of different letter doublets of codons is not large: 8 (one for each strong doublet) + 16 (two for each weak one) = 24. There is only one codon for which all three nucleotides are significant – UGA, the stop codon. UGA's potential anticodon – UCA, placed into the tryptophan tRNA, would not be able to encode tryptophan. It would keep the conformation of the first two nucleotides unchanged. This can be a way to verify my speculations experimentally.

**The triplet genetic code.** Speaking of the codon size, I suppose that it was originally triplet, even at the stage when there were just guanine and cytosine. In my opinion, this is not related to the necessity for four nucleotides to encode for twenty amino acids; it is rather the number of amino acids that is limited by the number of encoding variants of codon forms. Assuming that the code expanded incorporating new amino acids, it would be more natural to suggest that the number of amino acids should be limited by the capacity of the code rather than that the length of the encoding word should be determined by the number of amino acids.

The triplets represented by guanine and cytosine only cannot encode for more than four amino acids. This is in good agreement with my substantiation of Rumer's symmetry and the above suggestion that the form of the letter doublet is the recognizable component of the codon.

There is no need to speak of the doublet code, automatically applying Gamov's ideas to the early genetic code [22]. A simpler assumption is that the length of the codon has always been constant, and just its composition has been changed.

This assumption allows a natural explanation of why the replacement of purine by pyrimidine is extremely significant in the second letter and is often permissible in the first letter of the codon. Changes in the first letter of the triplet alter the shape of the doublet (the size of the first letter is altered), but this alteration affects the surroundings of the middle letter only. Changes in the second letter alter the surroundings of all letters, the whole shape of the codon. For the doublet code, changes would be equally significant, and insignificant variations in kindred codons would be out of the question. This does not give rise to any contradictions in the stage when there were just C, G, and U, because the only mutation that I dealt with in that stage was cytosine deamination. Cytosine deamination does not alter the size of the nucleotide. In that stage, however, I expect the emergence of the stop codon, and this can be more easily accounted for based on the triplet code hypothesis, as will be clear from the section below.

**Emergence of the stop codon. Part two.** I have already put forward the hypothesis that the stop codon emerged as a result of instability of interaction between the UG doublet and its potential anticodon. For this hypothesis it is essential to recognize the existence of the stage at which there is uracil but there is no adenine yet. As stated above, in this stage uracil is incorporated in the RNA double helix in its enol form. This is also valid for the codon-anticodon interaction, but specific properties of the micro-surroundings cannot be ignored. I suppose that the codon beginning with the UG doublet had originally been prone to have uracil in the keto form in its complementary interaction with the potential anticodon, which, respectively, began with GC.
What could be the difference between this codon and the other codons? There are grounds to suppose that its properties are related to the symmetry of strong and weak letter doublets discovered by Rumer. This was the only weak doublet that had emerged before adenine did.
Let us first look at the CU and GU doublets. If we assume the existence of the doublet code, uracil is at the edge and can be in its keto form. If the code is triplet, the small portion of uracil in the RNA of that stage makes G or C the most probable third letter. The enol form of uracil, restricted in this way, will be able to make at least two hydrogen bonds with the complementary guanine. This may be interpreted as an indirect proof of the code being triplet, but this indirect proof can be verified experimentally!
Let us now compare triplet codons with the UC and UG doublets. The UG doublet does not change its conformation in the contemporary genetic code and it seems that it was not prone to do this before. There is no evidence of uracil preferably being in the enol form, but, at the same time, nothing points to a greater stability of the keto form in the absence of adenine.
The UG doublet is very illustrative: it is both able to change its conformation and sensitive to impacts equal to the energy of one hydrogen bond. In this case, one hydrogen bond can come into being and cease to exist when uracil is transformed from the enol form to the keto one. To change the conformation, purine does not have to be replaced by pyrimidine in the third letter. The same force that was earlier (in discussing the tryptophan codon) applied at one side is now taken away at the other. The transformation of uracil into its keto form is stimulated by the inclination of this (triplet) codon towards conformational changes. This reasoning is certainly less forcible than experiment (quite feasible in this case), but more convincing than many quantum chemical calculations. Please note again that for this reasoning it is essential that the triplet code should have existed even when there were just C, G, and U.
If the CU, UC, and GU doublets are characterized by the presence of uracil in the enol form and the UG doublet – by the presence of uracil in the keto form, the loss of one (or even two) hydrogen bonds in the interaction with the respective anticodon can lead to the development of stop codon properties in a "natural" way. Again, these properties can be studied experimentally, in contrast to discussions concerning evolution of aminoacyl-tRNA synthetases.

**The first gene, the first protein.** With just the first four amino acids encoded by guanine and cytosine there is no chance for the synthesis of catalytically active proteins. Proline, glycine, alanine, and arginine – no combination of these amino acids can yield a wide diversity of structures and properties.
Moreover, I assume that there was no start signal or translation stop signal. One can only suppose that synthesis could start at any position. This assumption makes it equally probable for any codon to occur. Supposedly, the primary sequence of early genes was random, which would suggest the stability towards the reading-frame shift. This assumption cannot be regarded as sufficiently productive, as such a starting point of evolution is extremely different from any sequence of selective value.

Let us imagine the maximally ordered structure of the first gene, which is similar to the crystal structure. This structure would be advantageous for self-reproduction. The existing disordering factors (spontaneous mutations) will in the future cause the periodic crystal to be transformed to Schrödinger's "aperiodic" crystal [23].

The fact that it is equally probable to find any of these four amino acids in the protein suggests one assumption concerning the properties and structure of the first protein. The first protein could be either collagen-like protein. High proline content is a necessary condition for constructing collagen. It is the presence of proline in every third or fourth position that makes the formation of the triple helix possible.

An essential condition for collagen is strict periodicity of the primary structure. The triplet code and strict periodicity make the structure sensitive to the reading-frame shift. If the task is to determine the structure of the gene stable towards any reading-frame shift, this problem, no matter how surprising this may seem, has one rigorous solution:

 (CCGG)(CCGG)(CCGG)(CCGG)(CCGG)(CCGG) etc.

The primary protein sequence will consist of the repeated motif (ProAlaGlyArg).

This gene and this protein allow fulfilling the invariance condition for transposition of RNA strands and invariance to the reading-frame shift; moreover, this is the only solution.

Could there be any reasons for realizing this very gene in the form of the RNA double helix? Let us consider the extreme case of the RNA double helix, in which one strand is represented by polyguanine and the complementary one – by polycytosine. In their free states, in the solution, polycytosine and polyguanine can be considerably different from each other. The strong influence of stacking in polyguanine will result in the structure with a significantly extended helical turn. The formation of the double helix will be associated with an essential restructuring of the helix. With the polymer being sufficiently long, the kinetic barrier can be unsurpassable. The symmetric positions of cytosine and guanine in the complementary strands make their forms similar and, thus, not requiring large expenditures of energy and time for the formation of the double helix. Simple recurrence of the GC fragment is disadvantageous, because in this structure the influence of stacking is very low. The repeated motif (CCGG) may on the average lead to some energy benefit. To continue the logic of this reasoning, one can assume that the (CCCGGG) structure is even more stable, etc. As already stated in this section, polyguanine and polycytosine must have incompatible structures of RNA single helixes. What should be the length of oligomer to reveal this incompatibility? My substantiation of Rumer's symmetry of the genetic code table suggests that stacking considerably influences the shape of the RNA double helix segment formed by the codon and the anticodon. That is, even within one triplet, the length of the helical turn can be noticeably different in polycytosine and polyguanine. This difference is noticeable for the formation of the double helix.

In the proposed structure (CCGG) these differences are physically (or mathematically) impossible.

The idea that the evolution of protein biosynthesis started with a strictly periodic gene rather than with quasi-periodic structures (as proposed by Crick) can certainly be received skeptically. However, this initial state, perfectly stable towards the reading-frame shift has some advantages. There are genes that simultaneously encode for two proteins, as a result of the reading-frame shift. The existence of these genes contradicts the assumption of the initial existence of only such genes as are written without a reading-frame shift. Based on the evolution of all genes from the single initial state, which is perfectly stable to the reading-frame shift, one can study the models of the evolutionary emergence of genes encoding for more than one protein with a shift.

Another issue that may be approached in a way that is "automatically" determined by the presence of the poly(CCGG) structure as the first gene is the emergence of tRNA. The primary structure of tRNA must contain several complementary segments. The poly(CCGG) gene has the necessary properties intrinsically. The single-strand molecule of this RNA must form the secondary structure through the formation of double helix segments. An approach to the emergence of ribosomal RNAs can also be proposed: their secondary structure is also predetermined by the availability of significant complementary segments in the primary sequence. The existence of this first gene will imply the existence of the structures similar to the tRNA and ribosomal RNA even prior to their acquisition of their contemporary functions.

**Early biosynthesis of protein.**

Once I have made assumptions concerning the early genetic code and even the first gene and the first protein, I should propose a possible mechanism of protein biosynthesis. In my opinion, the most likely mechanism is direct biosynthesis of protein on the polynucleotide matrix, already proposed by Gamov in 1954 [22], in the first study of the genetic code.

Note that arginine has been demonstrated to be able to recognize its codon [13]. The other three earliest amino acids do not carry a positive charge and cannot be sorbed by the RNA polyanion so well. I can propose changing experimental conditions by supplementing the medium with metal cations, which can form complexes with RNA and amino acids. Magnesium and calcium cations can be the most suitable as they are ubiquitous. This can be a way to equalize the charge.

The above arguments suggest that the dinucleotide form is the meaningful part of the codon and this is valid for the whole table of the genetic code. This can be considered as an "atavism" of the epoch in which amino acids were directly recognized by codons. Then, what is the function of the third nucleotide? Although it did not take part in the recognition, it was most probably necessary for the amino acid-codon geometric correspondence – just a dimensional correspondence between the amino acid and the codon.

Gamov assumed that protein was synthesized immediately on the DNA double helix [22]. My assumption is immediate protein synthesis on the RNA double helix. Segments of the RNA double helix still work in ribosomes. Unlike DNA, the RNA double helix is rigid enough to set considerable segments into coordinated motion. This is very useful for catalysis.

Let us now address the mechanism of catalysis. Let us look at the RNA (DNA) double helix close to the "melting" point, where the double helix is untwined to make two single-strand molecules. This process is accompanied by a disruption of three hydrogen bonds in one base pair and binding of six molecules of water. This is a very strong dehydration effect.

The reaction of the formation of the peptide bond during protein synthesis is essentially dehydration reaction. Two amino acids lose one water molecule.

If amino acids have affinity for their codon, the formation of these complexes must induce melting of the double helix and subsequent dehydration. If two amino acids are joined to RNA next to each other, one water molecule can leave them and a peptide bond can be formed. Partial untwining of the double helix makes new RNA segments accessible and they are joined by new amino acids; thus, protein biosynthesis goes on.

The newly formed protein must be a mechanical obstacle to the restoration of the disrupted segment of the double helix and cannot be recognized by hydrogen bonds of nucleotides any more.

This mechanism makes biosynthesis possible without the specific catalyst moving along the RNA. It offers a probable dehydrating reagent. It rules out adaptors and reduces protein biosynthesis to the minimal number of interacting reagents.

Beginning with this state, the further evolution is potentially deducible.

Please note that guanine and cytosine have a number of advantages for this scheme: 1. The ability to form the RNA double helix. 2. The presence of a large number of hydrogen bonds for molecular recognition. 3. The availability of the best opportunities for binding water molecules. These advantages are all based on the possibility of forming three complementary hydrogen bonds.

**Acknowledgement**

The author would like to thank Krasova E. for her assistance in preparing this manuscript. I would like to thank Drs. Okladnikova N.N. and Boyandin A.N. for helpful discussions. I wish to express my gratitude to Dr. Wolfgang Pluhar for his appropriate suggestion and encourage.